\documentclass[12pt]{article}
\usepackage{latexsym}
\usepackage{epsfig}

\def\bg#1{\mbox{\boldmath$#1$}}

\newcommand{\del}{\partial}
\newcommand{\beq}{\begin{eqnarray}}
\newcommand{\eeq}{\end{eqnarray}}
\newcommand{\be}{\begin{eqnarray*}}
\newcommand{\ee}{\end{eqnarray*}}
\newcommand{\bk}{{\bf k}}
\newcommand{\bp}{{\bf p}}

\newcommand{\br}{{\bf r}}
\newcommand{\bx}{{\bf x}}

\newcommand{\e}{\epsilon}
\newcommand{\ve}{\varepsilon}

\newcommand{\om}{{\omega}}

\begin{document}

\centerline{\Large\bf{Effective electromagnetic theory for dielectric media}}
\bigskip
\centerline{ Finn Ravndal\footnote{On sabbatical leave of absence from  Institute of Physics, University of Oslo, N-0316 Oslo, Norway.}}
\bigskip
\centerline{\it Department of Physics, University of Miami, Coral Gables, FL 33124.}

\begin{abstract}

\small{Light in a dielectric medium moves slower than in vacuum. The corresponding electromagnetic field equations are then no longer invariant under ordinary Lorentz transformations, but only under such transformations corresponding to this reduced velocity. Based on this physical symmetry, an effective theory for low-energy electromagnetic phenomena in dielectrics is constructed. It has none of the problems of the old formulations of Minkowski and Abraham.  Dispersion in the optical regime and the Kerr effect arise in a natural way from higher order interaction terms. The effective field theory is quantized by standard methods and quantum corrections can be calculated in a systematic way. Thus it relates many different classical and quantum optical phenomena into a unified description.}

\end{abstract}


Light in a isotropic, dielectric medium with refractive index $n>1$ moves with a wave velocity $1/n$ when setting the velocity of light in vacuum to $c = 1$.  The Maxwell equations describing the electromagnetic field in matter are well known and one would think that a consistent, theoretical description of electrodynamics in dielectrics would exist.  But for close to a hundred years two different theories, one due Minkowski and the other to Abraham, have been used. They were both constructed to be consistent with the special theory of relativity in vacuum\cite{classics}. 

The main difference between the two formulations is found in the energy-moment\-um tensors. In the Minkowski theory this tensor is not symmetric and thus has problems with the conservation of angular momentum. Abraham instead imposed such a symmetry with the consequences that the momentum density of the electromagnetic field is reduced by the factor $1/n^2$ and a corresponding new volume force appears. When quantized, the photon in a medium has a non-vanishing four-momentum squared. Many attempts have been made to understand the underlying problems without any real success\cite{theory}. In  standard textbooks the Abraham formulations seem to be preferred\cite{books}\cite{LL}, while most experiments favor the Minkowski formulation\cite{exp}. However, for a few experiments where the system undergoes acceleration, the Abraham theory might be required\cite{Chiao}. 

In the following it is pointed out that covariance under vacuum Lorentz transformations is not consistent with the symmetries of the underlying Maxwell equations. Instead we propose an effective theory based on Lorentz invariance under transformations corresponding to the physical speed of light $1/n$ in the medium. We then avoid the old problems of Minkowski and Abraham. Since the medium itself is not invariant under these transformations, this new formulation is to be used in the rest frame of the medium. It is thus on an equal footing with other theories in condensed matter physics for excitations or quasiparticles with linear dispersion relations.  It is also shown how non-linear dispersion and the Kerr effects follow from higher-dimension operators in the effective theory.

The standard  theory for electromagnetic fields in a medium without any free currents or charges is given by the ordinary Maxwell equations ${\bg\nabla}\times{\bf E} + \del{\bf B}/\del t = 0$ and ${\bg\nabla}\times{\bf H} - \del{\bf D}/\del t = 0$ together with ${\bg\nabla}\cdot{\bf D} = {\bg\nabla}\cdot{\bf B} = 0$ where ${\bf D} = \ve {\bf E}$ and ${\bf B} = \mu {\bf H}$ for isotropic matter.  This is already an effective description on large scales where all the underlying atomic physics describing the material content of the medium is parametrized by the electric permittivity $\ve$ and the magnetic permeability $\mu$. We will use units so that in vacuum both $\ve_0 = 1$ and $\mu_0 = 1$ so that the speed of light $c = 1/\sqrt{\e_0\mu_0} = 1$ in vacuum. The electric and magnetic fields have then the same dimensions.

In the following both $\ve$ and $\mu$ are taken to be constants.  We thus first consider phenomena in the dielectric within a narrow range of frequencies where this assumption holds. The above Maxwell equations then define our field theory to lowest order. It is defined in the rest frame of the medium. Taking curl of the first Maxwell equation  and using the second equation, one then obtains in the usual way the wave equation
\beq
           \Big(n^2{\del^2\over\del t^2} - {\bg\nabla}^2\Big){\bf E}(\br,t) = 0                    \label{wave}
\eeq
for the electric field and the same for the magnetic field.  The electromagnetic wave velocity is therefore $1/n = 1/\sqrt{\ve\mu}$ in the medium. 

Neither this equation nor the original Maxwell equations are invariant under the ordinary Lorentz group. Instead they are invariant under what we  will call the material Lorentz group, with the vacuum light velocity replaced with $1/n$. The original treatments of electromagnetism in matter were generally formulated covariantly so to maintain invariance under the vacuum Lorentz group. We do not see how that can be justified for phenomena on a macroscopic scale where this does not represent an inherent symmetry of the system. It would not be  the first time in physics where the symmetry is different on different scales. 

It is straight forward to construct a special theory of relativity within the medium. One just has to replace the ordinary vacuum light velocity with the dielectric velocity $1/n$.  Taking the metric to be the same as in vacuum, a coordinate four-vector would then be $x^\mu =( t/n, \bx)$ so that the corresponding dielectric four-momentum becomes $p^\mu = (nE, \bp)$. The equivalent of a massless photon in the ordinary sense within the medium should therefore have energy and momentum related by the invariant $p^\mu p_\mu = 0$ or $E = p/n$. This is satisfied by the photon in the Minkowski theory. In solid state physics this is usually said to be an excitation or quasiparticle with a linear dispersion relation.

A covariant formulation based on this modified Lorentz group also follows naturally. Combining the electric potential $\Phi$ and the magnetic vector potential ${\bf A}$ into the four-vector potential $A^\mu = (n\Phi, {\bf A})$, the components of the electric field vector ${\bf E} = - {\bg\nabla}\Phi - \del{\bf A}/\del t$ and the ones of the magnetic field ${\bf B} = {\bg\nabla}\times{\bf A}$ form the antisymmetric tensor $F_{\mu\nu} = \del_\mu A_\nu - \del_\nu A_\mu$. The invariant Lagrangian density for the electromagnetic field within the dielectric is then
\beq
            {\cal L} = - {1\over 4\mu}F_{\alpha\beta}F^{\alpha\beta} = {1\over 2\mu}\Big(n^2{\bf E}^2 - {\bf B}^2\Big)                    \label{Lagrange}
\eeq
From here one finds the standard energy-momentum tensor in the medium to be given by $\mu T^{\mu}_{\;\;\nu} =  F^{\mu\alpha}F_{\alpha\nu} + ({1/4})\delta^{\mu}_{\;\;\nu}F_{\alpha\beta}^2$ which is obviously symmetric. Its conservation $\del_\mu T^{\mu}_{\;\;\nu} = 0$ follows from the Maxwell equations of motions. The time component of this conservation law gives ${\del{\cal E}/\del t} + {\bg\nabla}\cdot{\bf N} = 0$
where ${\cal E} = (1/2\mu)(n^2{\bf E}^2 + {\bf B}^2)$ is the energy density of the field and ${\bf N} = {\bf E}\times{\bf H}$ is the Poynting vector. This equation thus gives energy conservation. Similarly, the spatial components give the vector equation $\del{\bf G}/\del t + {\bg\nabla}\cdot{\bf T} = 0$ where the three-tensor ${\bf T}$ contains the Maxwell stress components $T_{ij}$. We therefore conclude that the vector ${\bf G} = {\bf D}\times{\bf B} =  n^2{\bf N}$ represents the momentum density of the field. 

Quantization of this free theory is straight forward. Integrating the above energy density, the resulting Hamiltonian gives a photon with wave number $\bk$ an energy  $\hbar\om_\bk$ where $\om_\bk = |\bk|/n$. Similarly, the momentum density ${\bf G}$ gives it a momentum $\hbar\bk$. The same energy and momentum also follow from the Minkowski theory. But while we would say that this particle in the medium is massless, in the Minkowski theory it has a negative mass squared, i.e. it is a tachyon.

There are two obvious experimental consequences one can extract right away from this free theory. One is the Casimir force\cite{Casimir-rev}, which now can be measured with good precision for the geometry of two parallel plates with separation $L$ in vacuum\cite{Casimir-exp}. Replacing the vacuum with a dielectric the force is expected to change. This experi\-mental situation has recently been considered by Brevik and Milton\cite{BM} who after a rather long calculation find within the Minkowski theory that the Casimir force should be reduced by a factor $n$ compared with the vacuum force $F = -\hbar\pi^2/240L^4$. This result now follows straight away from  the above since the force is essentially just the zero-point energy $E = \sum_\bk \hbar\om_\bk$ between the plates. The sum $E_0 = \sum_\bk \hbar |\bk |$ represents the corresponding zero-point energy in vacuum and is independent of the refractive index. It is not clear what the alternative Abraham theory will give for this Casimir force.

We can also make a similar, direct derivation of the energy density for black-body radiation in a cavity filled with a dielectric at temperature $T$. It follows from the usual expression
\be
       U = 2\sum_\bk{\hbar\om_\bk\over e^{\beta\hbar\om_\bk} - 1}
\ee
with $\beta = 1/k_B T$ where $k_B$ is the Boltzmann constant. Again with $\om = k/n$ this gives in the large-volume limit a result larger than the standard Stefan-Boltzmann formula by a factor $n^3$\cite{Lifshitz}. For a typical dielectric this increase will thus be in the range 2-7 and its effect should be detectable provided that this lowest-order theory contains the dominant physics.

Higher order corrections to these results can now be calculated within an effective field theory. Such theories have a long tradition in condensed matter physics and are now also used extensively in low- and high-energy particle physics. They are valid below an upper energy cut-off characteristic of the problem under investigation and involve only the relevant field variables of lower energy.  All degrees of freedom occurring on energy scales above this cut-off are integrated out to give higher dimensional couplings between the low-energy fields. Even if the theories are not renormalizable, quantum corrections can be calculated in a systematic way with divergences absorbed into the coupling constants. 

In order to be gauge invariant, higher order couplings can only involve the fields  ${\bf E}$ and  ${\bf B}$ and derivatives of them. For the sake of counting, we can assign the dimension +2 to each of these fields and an increase in dimension by +1 for every derivative using quantum units where $\hbar =1$. To be invariant under time-reversal and ordinary rotations, such new couplings must involve at least two space-time derivatives. For example, one possibility could be the term ${\bf E}\cdot\del^2{\bf E}$ with $\del^2 = \del^\mu\del_\mu = n^2\del_t^2 - \nabla^2$. It has dimension 6 and is also invariant under the material Lorentz group. But the lowest order equation of motion (\ref{wave}) is just $\del^2{\bf E} = 0$ and this term can therefore not contribute.  Possible new terms of dimension 8 would be $({\bf E}\cdot{\bf E})^2$,  $({\bf B }\cdot{\bf B})^2$,  ${\bf E}^2{\bf B}^2$ and $({\bf E}\cdot{\bf B})^2$. Requiring Lorentz invariance, the first three of these non-linear couplings must appear in the combination $(n^2{\bf E}^2 - {\bf B}^2)^2$. All such terms represent anharmonic interactions involving four fields. 

Dispersion can now be obtained from dimension-6 interactions when we break the material Lorentz invariance down to rotational invariance. One example is $\nabla_i{\bf E}\cdot\nabla_i{\bf E}$. It is equivalent to ${\bf E}\cdot\nabla^2{\bf E}$ by a partial integration in the action integral where it appears. The similar term  $\del_t{\bf E}\cdot\del_t{\bf E}$ involving two time derivatives is for the same reason equivalent to ${\bf E}\cdot\nabla^2{\bf E}$ when we use the equation of motion. An interaction like ${\bf E}\cdot\nabla^2{\bf B}$ is ruled out by parity invariance.

In dielectric materials magnetic effects are negligible, and it is therefore reasonable to assume that all the terms involving the magnetic field, can be dropped.  Including interaction terms up to dimension-8 operators, we then have the effective Lagrangian
\beq
           {\cal L} = {1\over 2}\Big(n^2{\bf E}^2 - {\bf B}^2\Big)  +  {d_1\over M^2} {\bf E}\cdot\nabla^2{\bf E}  
                       + {d_2\over M^4} ({\bf\nabla^2 E})^2  + {a\over M^4} ({\bf E }\cdot{\bf E})^2                                 \label{L_eff}
\eeq
The parameter $M$ is an energy scale which characterizes the microscopic physics we have integrated out. In a dielectric this is set by atomic physics and should have a value around 5 - 10 eV.  It is possible to calculate or at least estimate the dimensionless coupling constants $d_1$, $d_2$ and $a$ from the atomic physics of the material. With a realistic value for the energy scale $M$, the Lagrangian contains only three independent parameters. For each material they can therefore be determined by three different measurements. It would then be able to predict the outcome of many other experiments. In practice, we expect the term $\propto d_2$ to be dominated by the $d_1\,$-term.

Let us first consider the effect of the dispersive term $\propto d_1$ in the effective Lagrangian. It gives a contribution to the classical equation of motion  for the field that is linear. For a plane wave of the form ${\bf E} \propto e^{i(\bk\cdot\bx - \om t)}$ we obtain a dispersion given by $\om^2 = k^2/n^2 + 2d_1\om^2k^2/\ve M^2$. Since the last term is assumed to be small, we can there set $\om  = k/n$.  With $\ve = n^2$  in a dielectric,  we find the refractive index, defined by the phase velocity $\om/k \equiv 1/n(\om)$, to be $n(\om) = n(1 - d_1\om^2/M^2)$. This is of the same form as the phenomenological law due to Cauchy and usually written as $n(\lambda) = A + B/\lambda^2$\cite{AN}. The measured values of the dispersion parameter $B$ for most gases and transparent dielectrics are in the rather narrow range $(2-15)\times 10^{-15}\,\mbox{m}^2$. The coupling constant $d_1$ must therefore be negative to give normal dispersion. Comparing with our result, we find that these measured values correspond to taking  $nM = 10\,\mbox{eV}$ and -$d_1$ in the range $0.1 -  1.0$. This is just as expected from the scale of the underlying atomic physics .  Including the $d_2\,$-term, we would similarly have have obtained a $\om^4$ correction to the above dispersion relation.

The four-field coupling in (\ref{L_eff}) can describe many non-linear phenomena in dielectrics. It has a coupling constant $a$ which corresponds to the third-order susceptibility $\chi^{(3)}$ in the language of non-linear optics\cite{Boyd}. In the DC Kerr effect an external, electric field $E$ affects the propagation of light. The refractive index becomes field dependent. Here it results by replacing one factor ${\bf E }\cdot{\bf E}$ in the interaction term with the square $E^2$ of the external field. The remaining, oscillating field then obeys a linear equation of motion as above giving the refractive index $n(E) = n + 2aE^2/nM^4$.  Measurements are often parametrized as $n(E) = n + \lambda KE^2$  where $\lambda$ is the wavelength. Here $K$ is the Kerr constant which we thus find to be given by $\lambda K = 2a/nM^4$. For isotropic glasses it has values of the order $10^{-16}\,\mbox{m}/\mbox{V}^2$  corresponding to a coupling constant $a \approx 10^{-6}$ when  we take $nM = 10\,\mbox{eV}$.  For dielectric liquids this effect is usually larger.

In the AC Kerr effect the field of the wave itself affects its propagation.  Using a semi-classical approximation, we can again replace  one factor ${\bf E }\cdot{\bf E}$ in the interaction with the intensity  $I = \e E^2$ of the beam. Then we find for the refraction index $n(I) = n(1 + aI/(nM)^4)$. Writing it on the  form $n(I) = n + n_2I$, we thus have $n_2 = a/n^3M^4$. Measurements in optical glasses\cite{Boyd} give values for $n_2$ of the order $10^{-20}\,\mbox{m}^2/\mbox{W}$. With $nM = 10\,\mbox{eV}$ this is consistent with $a\approx 10^{-6}$ as obtained above. In fact, eliminating $a$ between these two Kerr constants, we find $\lambda K = 2n^2n_2$. This relation seems to be roughly satisfied by data both for optical glasses and for dielectric liquids. A similar relation can also be derived by simple, atomic arguments\cite{Boyd}.

With this effective field theory we can now also calculate quantum corrections to some of the processes previously discussed.
For the energy of black-body radiation in a dielectric the calculation would be very similar to what has already been done\cite{KR-1} based upon the corresponding effective Euler-Heisenberg theory for low-energy QED\cite{EH}.  In that theory the electron mass $m_e$ plays the role of $M$ here and the dimensionless coupling constants are powers of the fine-structure constants and thus very small.  Since $m_e/M \approx 10^6$, the higher order corrections we consider in a material  would therefore be enormously bigger than these QED effects and therefore in principle  experimentally more accessible. But here the dominant contribution will come from the $d_1$ dispersive term. The corresponding term does not exist in the Euler-Heisenberg effective Lagrangian because that theory is required to be Lorentz invariant.  It will give a correction with size of the order $(k_BT/M)^2$ times the lowest-order result. Taking $M = 10$\,eV,  this would first start to be signi\-ficant at such high temperatures that the material has already melted. But here the lowest-order result itself is important to verify experimentally.

Corrections to the  Casimir force can also be obtained analogously to calculations previously done in vacuum\cite{KR-2}, but now based on the more important $d_1$ dispersive term. However, for the Casimir force it is possible that a surface term will also  be present in the effective theory\cite{RT}\cite{AB}. The dominant correction will then be of the order $1/LM$ times the vacuum force, provided that the scale of this interaction is of the same order of magnitude as in the bulk.  It will then begin to be important for plate separations approaching the cut-off distance $L = 10\,\mbox{nm}$. With modern nanotechnology this should soon be within reach of experiments\cite{Onofrio}. For even shorter separations the effective theory is not expected to be valid anymore.
 
Looking back at previous theories, one sees at least two reasons for their difficulties. The first is that they were covariantly formulated based on the vacuum Lorentz group, a symmetry which is not present in the macroscopic Maxwell equations. In the present proposal the rest frame of the medium is a preferred frame. Secondly, the problem with quantizing them was frustrated by the effect of dispersion.  Here dispersion is not part of the lowest-order theory, but arise as a higher order effect which is included using ordinary perturbation theory.

We can also write down corresponding effective Lagrangians for electromagnetic fields in anisotropic materials. The rotational invariance so far assumed will then be replaced with the invariance under a discrete, crystal group appropriate for that particular material. As a consequence, the coupling constants appearing in (\ref{L_eff}) will be tensors and new couplings can arise. Needless to say, there will then be a few more unknown parameters,  but the effective theory might still be useful.

This work has been helped by discussions with I. Brevik, C. Burgess, T. Curtright and L. Mezincescu.  Colleagues at the Department of Physics at UM are acknowledged for their hospitality.


\begin{thebibliography}{99}
\bibitem{Lifshitz} The same result is derived in the textbook \cite{LL} by Landau and Lifshitz based on the correlators of fluctuating currents in in the cavity walls. But they notice at the end of the calculation that the result can be obtained more directly from the relation $\om = k/n$. It is not clear what the Abraham formalism they espouse elsewhere, would give.
\end{thebibliography}
\end{document}